\documentclass[10pt,letterpaper,conference]{ieeeconf}

\IEEEoverridecommandlockouts
\usepackage{graphicx}
\usepackage{color}
\usepackage[cmex10]{amsmath}
\usepackage{amssymb}
\usepackage[tight,footnotesize]{subfigure}
\usepackage{float}
\usepackage{multirow}
\usepackage{booktabs}
\usepackage{subeqnarray}
\usepackage{enumerate}
\usepackage{epstopdf}
\usepackage{wrapfig}
\usepackage{soul}
\usepackage[table]{xcolor}
\usepackage{array}
\usepackage[ruled,linesnumbered]{algorithm2e}

\usepackage{cite}
\usepackage[bookmarks=true,colorlinks=true,linkcolor=black,backref=true,
pdfborder={0 0 0},citecolor=blue,urlcolor=blue]{hyperref}

\makeatletter
\let\proof\@undefined			
\let\endproof\@undefined		
\makeatother
\usepackage{amsthm}

\usepackage{tikz}
\usetikzlibrary{shapes.geometric, arrows, positioning, arrows.meta}
\tikzstyle{process} = [rectangle, rounded corners, minimum width=1cm, minimum height=0.75cm, 
text centered, draw=black, fill=white]
\tikzstyle{arrow} = [line width=1mm,->,>=stealth, blue!20]

\usepackage[thinc]{esdiff}

\theoremstyle{definition}

\allowdisplaybreaks

\newcommand{\real}[1][]{\mathbb{R}^{#1}}                                
\newcommand{\nat}[1][]{\mathbb{N}^{#1}}                                 

\newcommand{\defeq}{:=}                                                 
\newcommand{\msub}[1]{_\mathrm{#1}}                                     

\newcommand{\myfracB}[2]{{#1}/{#2}}                                     

\newcommand{\mydiff}[2]{\frac{\mathrm{d}{#1}}{\mathrm{d}{#2}}}      	
\newcommand{\df}{\mathrm{d}}                                            
\newcommand{\parder}[2]{\frac{\partial #1}{\partial #2}}                
\newcommand{\parders}[2]{{#1}_{#2}}   

\newcommand{\atan}[1]{\tan^{-1}\left(#1\right)}                         

\renewcommand{\geq}{\geqslant}                                          

\newcommand{\mydef}[1]{{\textit{#1}}}

\newcommand{\figpath}{Figures}

\newcommand{\eqnnt}[1]{\hyperref[#1]{(\ref*{#1})}}
\newcommand{\eqnsnt}[2]{\hyperref[#1]{(\ref*{#1})}
	and~\hyperref[#2]{(\ref*{#2})}}
\newcommand{\eqnsernt}[2]{\hyperref[#1]{(\ref*{#1})}--\hyperref[#2]{(\ref*{#2})}}

\newcommand{\eqn}[1]{\hyperref[#1]{Eqn.~(\ref*{#1})}}
\newcommand{\eqns}[2]{\hyperref[#1]{Eqns.~(\ref*{#1})} and~\hyperref[#2]{(\ref*{#2})}}
\newcommand{\eqnser}[2]{\hyperref[#1]{Eqns.~(\ref*{#1})}--\hyperref[#2]{(\ref*{#2})}}
\newcommand{\eqnf}[1]{\hyperref[#1]{Equation~(\ref*{#1})}}
\newcommand{\eqnfs}[2]{\hyperref[#1]{Equations~(\ref*{#1})} and~\hyperref[#2]{(\ref*{#2})}}

\newcommand{\scn}[1]{\hyperref[#1]{Sec.~\ref*{#1}}}
\newcommand{\scns}[2]{\hyperref[#1]{Secs.~\ref*{#1}} and~\hyperref[#2]{\ref*{#2}}}
\newcommand{\scnser}[2]{\hyperref[#1]{Secs~.\ref*{#1}}--\hyperref[#2]{\ref*{#2}}}

\newcommand{\fig}[1]{\hyperref[#1]{Fig.~\ref*{#1}}}
\newcommand{\figs}[2]{\hyperref[#1]{Figs.~\ref*{#1}} and~\hyperref[#2]{\ref*{#2}}}
\newcommand{\figser}[2]{\hyperref[#1]{Figs.~\ref*{#1}}--\hyperref[#2]{\ref*{#2}}}
\newcommand{\figf}[1]{\hyperref[#1]{Figure~\ref*{#1}}}
\newcommand{\figfs}[2]{\hyperref[#1]{Figures~\ref*{#1}} and~\hyperref[#2]{\ref*{#2}}}
\newcommand{\figfser}[2]{\hyperref[#1]{Figures~\ref*{#1}}--\hyperref[#2]{\ref*{#2}}}

\newcommand{\tbl}[1]{\hyperref[#1]{Table~\ref*{#1}}}
\newcommand{\tbls}[2]{\hyperref[#1]{Tables~\ref*{#1}} and~\hyperref[#2]{\ref*{#2}}}
\newcommand{\tblser}[2]{\hyperref[#1]{Tables~\ref*{#1}}--\hyperref[#2]{\ref*{#2}}}

\newcommand{\apx}[1]{\hyperref[#1]{Appendix~\ref*{#1}}}

\newcommand{\prb}[1]{\hyperref[#1]{Problem~\ref*{#1}}}
\newcommand{\prp}[1]{\hyperref[#1]{Prop.~\ref*{#1}}}
\newcommand{\prpf}[1]{\hyperref[#1]{Proposition~\ref*{#1}}}

\newcommand{\algoref}[1]{\hyperref[#1]{Algorithm~\ref*{#1}}}

\newcommand{\thmref}[1]{\hyperref[#1]{Theorem~\ref*{#1}}}
\newcommand{\thmsref}[2]{\hyperref[#1]{Theorems~\ref*{#1}} and~\hyperref[#2]{\ref*{#2}}}
\newcommand{\thmserref}[2]{\hyperref[#1]{Theorems~\ref*{#1}}--\hyperref[#2]{\ref*{#2}}}

\newcommand{\algline}[1]{\hyperref[#1]{Line~\ref*{#1}}}
\newcommand{\alglines}[2]{\hyperref[#1]{Lines~\ref*{#1}} and~\hyperref[#2]{\ref*{#2}}}
\newcommand{\alglineser}[2]{\hyperref[#1]{Lines~\ref*{#1}}--\hyperref[#2]{\ref*{#2}}}

\renewcommand{\vec}[1]{\boldsymbol{#1}}
\def\threat{c}
\def\x{\vec{x}}
\def\p{\vec{p}}

\def\wsp{\mathcal{W}}

\def\pos{\vec{r}}
\def\posx{r_x}
\def\posy{r_y}

\def\dposx{\dot{r}_x}
\def\dposy{\dot{r}_y}

\def\pos{\vec{x}}
\def\dpos{\dot{\vec{x}}}
\def\posx{x_1}
\def\posy{x_2}

\def\dposx{\dot{x}_1}
\def\dposy{\dot{x}_2}

\def\cost{J}

\def\ui{\hdng}

\def\spd{v}

\def\hdng{\psi}
\def\tf{t\msub{f}}
\def\htf{\hat{t}\msub{f}}
\def\constA{\lambda}

\def\xInit{\x_0}
\def\xFin{\x\msub{f}}
\def\nParam{N\msub{P}}

\newcommand{\hamiltonian}{H}

\newcommand{\nnParam}{\theta}
\newcommand{\nnParamWt}[1][]{\nnParam_{\mathrm{w}#1}}
\newcommand{\nnParamBs}[1][]{\nnParam_{\mathrm{b}#1}}

\newcommand{\datum}{x}
\newcommand{\nData}{N\msub{D}}
\newcommand{\latentVector}{z}
\newcommand{\yOutput}{y}

\newcommand{\Loss}{\mathcal{L}}
\def\BibTeX{{\rm B\kern-.05em{\sc i\kern-.025em b}\kern-.08em
		T\kern-.1667em\lower.7ex\hbox{E}\kern-.125emX}}
\usepackage{xfrac}

\title{\Large\bf Trajectory Optimization for Minimum Threat Exposure \\
using Physics-Informed Neural Networks}
\author{Alexandra E. Ballentine$^{\ast}$ 
	and Raghvendra V. Cowlagi$^{\ast\dagger}$   
\thanks{$^{\ast}$Aerospace Engineering Department, 
	Worcester Polytechnic Institute, Worcester, MA, 
	USA.\quad $^{\dagger}$Corresponding author. \quad
	Email: 
	\texttt{aeballentine, rvcowlagi@wpi.edu}}
}

\def\figpath{Figures}

\begin{document}
\maketitle
\begin{abstract}
We apply a physics-informed neural network (PINN)
to solve the two-point boundary value problem (BVP) arising from the
necessary conditions postulated by Pontryagin's Minimum Principle for
optimal control. Such BVPs are known to be numerically difficult to
solve by traditional shooting methods due to extremely high
sensitivity to initial guesses. In the light of recent successes in
applying PINNs for solving high-dimensional differential equations,
we develop a PINN to solve the problem of finding trajectories with
minimum exposure to a spatiotemporal threat for a vehicle kinematic
model. First, we implement PINNs that are trained to solve the BVP
for a given pair of initial and final states for a given threat
field. Next,  we implement a PINN conditioned on the initial state
for a given threat field, which eliminates the need for retraining
for each initial state. We demonstrate that the PINN outputs satisfy
the necessary conditions with low numerical error.
\end{abstract}

\section{Introduction}

Pontryagin's Minimum Principle (PMP) provides first-order necessary
conditions that must be satisfied by optimal control inputs of
dynamical systems. The evaluation of these conditions has been the
bedrock of optimal control solutions to many important problems in
aerospace engineering \cite{Bryson1975} such as many variants of the
Zermelo minimum-time navigation problem in drift fields, the Dubins
problem of kinematically constrained minimum-time navigation,
spacecraft orbit transfer, reentry, and landing.

The PMP necessary conditions result in a set of ordinary differential
equations in the state and co-state. Transversality conditions in the
PMP provide boundary conditions. In some problems,
one may be able to draw conclusions about the optimal
control through analyses specific to the problem. In general, however,
one must resort to numerical solutions of the boundary value problem
(BVP) resulting from the PMP to find the optimal control input.

Solutions to these BVPs are notoriously difficult because of extremely
high sensitivity to initial guesses. The co-state equations can be
unstable even if the original dynamical system is stable. Furthermore,
the boundary conditions are often \emph{split} due to the
transversality conditions, resulting in a two-point BVP (TPBVP), e.g.,
if the state is fixed at a boundary, then the co-state is free. These
difficulties often preclude the application of PMP necessary
conditions for designing optimal control inputs in many practical
problems, especially when real-time computation is desired.

Physics-informed neural networks (PINNs) have
revolutionized computational solutions of ordinary- and partial
differential equations~\cite{raissi2019physics,karniadakis2021physics}. 
Briefly, PINNs exploit the differentiability
of neural networks and use automatic differentiation to embed
residuals of differential equations in their loss functions. Over a
short period of time, PINNs have been applied for a wide variety of
problems including forward and inverse problems in fluid dynamics
\cite{cai2021physics, jagtap2022physics},
solid mechanics \cite{bai2023physics}, 
heat transfer \cite{cai2021physics}, power systems \cite{Huang2023}, 
and finance \cite{nuugulu2024physics}.

The literature reports PINNs for solving optimization and optimal 
control problems \cite{seo2024solving}. One category of works reports
direct optimization of the initial and boundary conditions of partial 
differential equations, e.g., \cite{MOWLAVI2023111731,Lai2025}.
Another category reports 
model-based reinforcement learning (RL), where PINNs are reported
for efficient adaptive sampling or system identification in
low-data regimes, e.g., 
\cite{pmlr-v211-ramesh23a,
	BANERJEE2025128166,SAEED2024114879,Nagel2024}.
Solutions of the Hamilton-Jacobi-Bellman (HJB)
equation to find an optimal feedback policy in special cases,
e.g., \cite{Fotiadis2025,Nakamura2021}, 
as well as RL with inductive bias based on the HJB
\cite{mukherjee2023bridgingphysicsinformedneuralnetworks}
are reported.

In this paper, we apply PINNs for solving a BVP
resulting from PMP necessary conditions. The optimal control problem
of interest is to finding trajectories with minimum exposure to a
threat for a vehicle kinematic model. The threat is a spatiotemporally
varying scalar field, whose values and gradients are known.
In all but the most simple cases, such as a threat
field with a constant gradient, this problem does not present
analytical solutions. When the threat field is time-invariant but has
otherwise no special spatial properties, the co-states can be
eliminated via analysis and the solution is reduced to a tractable
BVP in the state variables and control input. Nevertheless, this BVP does
not present analytical solutions, and the initial control input value
must be determined to match the problem's given boundary conditions.
For time-varying threat field case, the co-states cannot be eliminated.

The main contribution of this work is a demonstration that PINNs
can provide a viable numerical method of solving TPBVPs for optimal
control problems that are challenging and impractical to solve using
conventional discretization-based methods. To this end, we implement a
PINN to solve the BVP and demonstrate accurate solutions to the
minimum threat exposure problem. First, we implement PINNs that are
trained to solve the BVP for a given pair of initial and final states
for a given threat field. The training process itself is agnostic to
these quantities. Next,  we implement a PINN conditioned on the
initial state for a given threat field, which eliminates the need for
retraining for each initial state. Through numerical studies we 
demonstrate that the PINN outputs satisfy the necessary conditions
with low error.


In what follows,
we assume that the reader is familiar with applied variational 
optimal control theory and Pontryagin's Minimum Principle, 
e.g.,\cite{Bryson1975}.

\section{Problem Formulation}
\label{sec-problem}

Consider a compact two-dimensional (2D) workspace $\wsp \subset
\real[2]$ attached with a Cartesian coordinate axes system, which we
assume is inertial. We denote by $\pos = (\posx, \posy)$ the position
coordinates of any location in this workspace. Let $\threat: \wsp
\times \real_{\geq 0} \rightarrow \real_{\geq 0}$ denote a
spatiotemporal scalar field that we call the \mydef{threat field}. The
objective is to find a trajectory with minimum threat exposure for a
vehicle moving from an initial location $\xInit \in \wsp$ to a final
location $\xFin \in \wsp.$

Consider the 2D vehicle kinematic model
\begin{align}
	\dposx(t) &= \spd \cos{\hdng(t)}, &
	\dposy(t) &= \spd \sin{\hdng(t)},
	\label{eq-kinematics}
\end{align}
where $\spd$ and $\hdng$ are the magnitude and direction
of the velocity vector, respectively. We assume $\spd$
fixed to a non-zero and known constant\footnote{If $\spd$ is 
a variable and upper-bounded, a consequence of the PMP for this
problem is that the optimal speed is always at the upper 
bound \cite{Cowlagi-GNC2015}.}. The heading angle $\hdng$ is
the control input.

The Bolza formulation of the minimum threat exposure problem
is to minimize the cost functional
\begin{align}
	\cost[\ui]  &\defeq \int_{t_0}^{\tf} \left(\constA + 
	\threat(\pos(t), t) \vphantom{1^2}\right) \df t,
\end{align}
where $\constA \geq 0$ is a prespecified constant, $t_0 \geq 0$ is
prespecified, and $\tf > t_0$ is free (i.e., to be determined).
Per the PMP, we write the Hamiltonian as
\begin{align}
	\hamiltonian(\x, \ui, \p, t) &= \constA + 
	\threat(\x, t) + p_1 \spd \cos\hdng
	+ p_2 \spd \sin\hdng,
	\label{eq-min-threat-H}
\end{align}
where $\p = (p_1, p_2)$ is the co-state.
The necessary conditions for an optimal trajectory
$\pos^*, \p^*, \hdng^*$ are:
\begin{align}
	\dpos^*(t) &= \parders{\hamiltonian}{\p}
	(\x^*, \ui^*, \p^*, t), \label{eq-pmp-state} \\
	\dot{\p}^*(t) &= -\parders{\hamiltonian}{\x}
	(\x^*, \ui^*, \p^*, t), \label{eq-pmp-co-state} \\
	\ui^*(t) &= \arg \min \left\{ 
	\hamiltonian(\x^*, \ui, \p^*, t) \right\},
	\label{eq-pmp-control}
\end{align}
along with the transversality condition
\begin{align}
	\hamiltonian(\x^*, \ui^*, \p^*, \tf) &= 0.
	\label{eq-min-threat-Htf}
\end{align}
Due to the boundary conditions $\x(0) = \xInit$
and $\x(\tf) = \xFin,$ the co-states $\p(0)$
and $\p(\tf)$ are free. 
Equations~(\ref{eq-pmp-state})-(\ref{eq-pmp-co-state}) lead~to
\begin{align}
	\dposx^*(t) &= \spd \cos{\hdng^*(t)}, &
	\dposy^*(t) &= \spd \sin{\hdng^*(t)}, 
	\label{eq-optimal-state} \\
	\dot{p}_1^*(t) &= -\parders{\threat}{\posx}(\x^*(t), t), &
	\dot{p}_2^*(t) &= -\parders{\threat}{\posy}(\x^*(t), t),
	\label{eq-optimal-co-state}
\end{align}
whereas \eqnnt{eq-pmp-control}, via $\parders{\hamiltonian}{\hdng} = 0,$ 
leads to
\begin{align}
	\hdng^*(t) &= \atan{\myfracB{p_2^*(t)}{p_1^*(t)}}.
	\label{eq-optimal-control}
\end{align}

The main problem of interest in this paper is to solve
the two-point BVP defined by
\eqnser{eq-min-threat-Htf}{eq-optimal-control}.

\section{Method of Solution}
\label{sec-method}

Equations~\eqnsernt{eq-optimal-state}{eq-optimal-control}, in general,
do not lend themselves to an analytic solution. In the special case
where the threat field is time-invariant, further simplifications may be made.
Note from the PMP necessary conditions that $\mydiff{\hamiltonian}{t} =
\parder{\hamiltonian}{t}$ along the optimal trajectory. 
Due to \eqnnt{eq-min-threat-H},
\begin{align}
	\mydiff{\hamiltonian}{t} (\x^*, \ui^*, \p^*, t) &= 
	\parder{\threat}{t} (\x^*, t),
	\label{eq-hamiltonian-derivative}
\end{align}
In the case where $\threat = \threat(\x^*)$, i.e. $\threat$ does not depend on $t$, 
$\parder{\threat}{t} =0$ and $\hamiltonian(\x^*, \psi^*, \p^*, t) = 0$ according to \eqnnt{eq-min-threat-Htf}.

Numerical solutions of \eqnser{eq-optimal-state}{eq-optimal-control}
by the traditional approach of a shooting method using 
a Runge-Kutta solver often fail to converge due to sensitivity 
to guesses of the initial co-state $\p(0).$ Even it
does converge, the speed of computation can be too slow for real-time
applications. To mitigate these issues, we seek to develop 
and evaluate a PINN solver for this BVP.

Artificial neural networks (NN) are universal function
approximators~\cite{LESHNO1993861}. Briefly, a
single-layer NN may be considered a nonlinear function of the form
$f(\datum; \nnParam) = \sigma(\nnParamWt^\intercal \datum +
\nnParamBs),$ where $\datum$ is the input, $\nnParam = (\nnParamWt,
\nnParamBs)$ are parameters consisting of weights $\nnParamWt$ and
biases $\nnParamBs,$ and $\sigma$ is a nonlinear activation function
such as the sigmoid function. By extension, a multi-layer or deep NN
may be considered a sequential composition of nonlinear functions of
the form $f(\datum; \nnParam) = \sigma(\nnParamWt[d]^\intercal
\latentVector_{d-1}+ \nnParamBs[d]),$ where $d \in \nat$ is the number
of layers, $\latentVector_{1} \defeq \sigma(\nnParamWt[1]^\intercal
\datum + \nnParamBs[1]),$ and $\latentVector_{k} \defeq
\sigma(\nnParamWt[k]^\intercal \latentVector_{k -1} + \nnParamBs[k])$
for $k = 2, \ldots, d-1.$ The neural network \mydef{learns} or
\mydef{is trained} over a dataset of input-output pairs $\{(\datum^i,
\yOutput^i)\}_{i=1}^{\nData}.$ Training is accomplished by finding
parameters $\nnParam^*$ that minimize a \mydef{loss function} $\Loss$, i.e.,:
$\nnParam^* \defeq \arg \min_{\nnParam} \Loss (\datum, \yOutput, \nnParam).$

%
%
%

The exact form of the loss function depends on the application. 
A common example is the mean square loss function $\Loss (\datum,
\yOutput, \nnParam) \defeq \frac{1}{\nData} \sum_{i=1}^{\nData} \|
\yOutput^i - f(\datum^i; \nnParam) \|^2.$ 
A distinctive property of a
PINN is that the loss function involves the residual of some governing
equation that the input-output pair $\datum, \yOutput$ must satisfy.
For example, suppose the differential equation $\mydiff{\yOutput}{\datum} = a \datum$
governs $\datum$ and $\yOutput.$ Then the loss function may include
the residual term $\| \mydiff{f(\datum; \nnParam)}{\datum} 
- a \datum\|$. 
Crucially, the exact computation of the derivative
$\mydiff{f(\datum; \nnParam)}{\datum}$ is enabled by automatic
differentiation \cite{baydin2018automatic}. The interested reader is
referred to \cite{raissi2019physics} for more detail 
on PINNs and their applications for solving
ordinary and partial differential equations.

We design a pair of neural networks that both accept initial state
$\x_0 \in \real[2]$ and normalized time $\tau \in [0, 1]$ as input. 
The outputs of the first network are a state  $\hat{\x} =
(\hat{x}_1, \hat{x}_2)$ and control
input $\hat{\hdng}$ at $\tau$, and the anticipated final time $\htf.$ The outputs of the second network are 
the co-states $\hat{\p}=(\hat{p}_1, \hat{p}_2)$ at $\tau$.
We denote the NN predictions with a hat, e.g., $\hat{\x}.$

The final location of the agent and the threat field are implicitly
learned by the PINN. We choose to
keep these variables implicit to the network to seek a
high-fidelity solution to a simple problem.

\subsection{Loss Functions}

Due to the number of constraints on the system, we define multiple
loss functions. Note given $\tau=\frac{t}{t_f}$ and a function $y=y(x, \tau)$, $\frac{\partial y}{\partial t} = \frac{\partial y}{\partial \tau}\mydiff{\tau}{t}=\frac{1}{t_f}\frac{\partial y}{\partial \tau}.$ 

The first loss function $\Loss_1$ relates the desired value of the Hamiltonian $\hamiltonian_d$ to $\hat{\hamiltonian}$ at $N$ collocation points:
\begin{equation*}
	\Loss_1 \defeq \frac{1}{N} \sum_{i=1}^N \bigl| 
	\hamiltonian_d(\tau) - (\hat{p}_1^i \spd \cos \hat{\psi}^i + 
	\hat{p}_2^i \spd \sin \hat{\psi^i} + \threat^i) \bigl|^2,
\end{equation*}
where index $i$ represents the network output for input $\tau$. For a static threat field, $H_d = 0$. For a dynamic threat field, we determine $H_d$ by backward integrating $\mydiff{H}{t}$ from the boundary condition in \eqnnt{eq-min-threat-Htf} using a midpoint approximation:

\begin{align*}
	\hamiltonian_d(\tau) = -\htf \sum_{j=N-1}^{i} \frac{1}{2}(\parder{{c}^{j+1}}{t} + \parder{{c}^{j}}{t}) \Delta \tau,
\end{align*}
where $\parder{{c}}{t}$ is the derivative of the threat field w.r.t. time. $j$ indexes the trajectory, and $i$ corresponds to the index of $\tau$.

The second loss function $\Loss_2$ describes the time derivative of the Hamiltonian:
\begin{equation*}
	\Loss_{2} \defeq 
	\frac{1}{N} \sum_{i=1}^{N} \bigl| \diffp{c^i}{t} - 
	\diff{}{t}(\hat{p}_1^i \spd \cos \hat{\psi}^i + 
	\hat{p}_2^i \spd \sin \hat{\psi}^i + c^i) \bigl|^2,
\end{equation*} 
where $\parder{c^i}{t}$ is the analytic time derivative of the threat field at the current state.

We define
\begin{align*}
	\Loss_{3} &\defeq \frac{1}{N} \sum_{i=1}^N \bigl| \frac{1}{\htf} \diff{\hat{x}_1^i}{\tau} -
	\spd \cos \hat{\psi^i} \bigl|^2, \\
	\Loss_{4} &\defeq \frac{1}{N} \sum_{i=1}^N \bigl| \frac{1}{\htf} \diff{\hat{x}_2^i}{\tau} - 
	\spd \sin \hat{\psi}^i \bigl|^2,
\end{align*}
The derivatives
$\mydiff{}{\tau}$ are computed 
w.r.t. normalized time using automatic differentiation.
We define losses based on \eqnnt{eq-optimal-co-state}:
\begin{align*}
	\Loss_{5} &\defeq \frac{1}{N} \sum_{i=1}^N \bigl| 
	\frac{1}{\htf} \diff{ \hat{p}_1^i }{\tau} + 
	\diffp{c^i}{ {{\hat{x}_1^i}} } \bigl|^2, \\
	\Loss_{6} &\defeq \frac{1}{N} \sum_{i=1}^N \bigl| 
	\frac{1}{\htf}  \diff{\hat{p}_2^i}{\tau} + \diffp{c^i}{{{\hat{x}_2^i}}} \bigl|^2.
\end{align*}
We define loss $\Loss_7$ based on \eqnnt{eq-optimal-control} to encode the relationship between the co-states and the optimal control input.
\begin{equation*}
	\Loss_7 \defeq \frac{1}{N} \sum_{i=1}^N \bigl| 
	- \hat{p}_1^i \spd \sin \hat{\psi}^i 
	+ \hat{p}_2^i \spd \cos \hat{\psi}^i\bigl|^2.
\end{equation*}

Next, we consider the boundary conditions
$\x(0) = \xInit$ and $\x(\tf) = \xFin.$
We define two loss functions based on the errors
between the boundary states predicted by the PINN
against the desired states:
\begin{align*}
	\Loss_8 &\defeq \| \hat{\x}_0 - \x(0) \|^2, & 
	\Loss_9 &\defeq \| \hat{\x}\msub{f} - \xFin \|^2.
\end{align*}

Finally, we define a loss $\Loss_{10}$ to penalize high cost paths. 
Because the PMP conditions are \emph{necessary} but not
\emph{sufficient}, finding a candidate trajectory that satisfies the
PMP conditions does not guarantee it is a minimum. 
We use a right Riemann sum to approximate the integral path cost, i.e. 
$\Loss_{10} \defeq \htf \sum_{i=2}^N
	c(\hat{\x}^i, \tau^i) \Delta \tau.$

\begin{figure*}
	\centering
	\includegraphics[width=\linewidth]{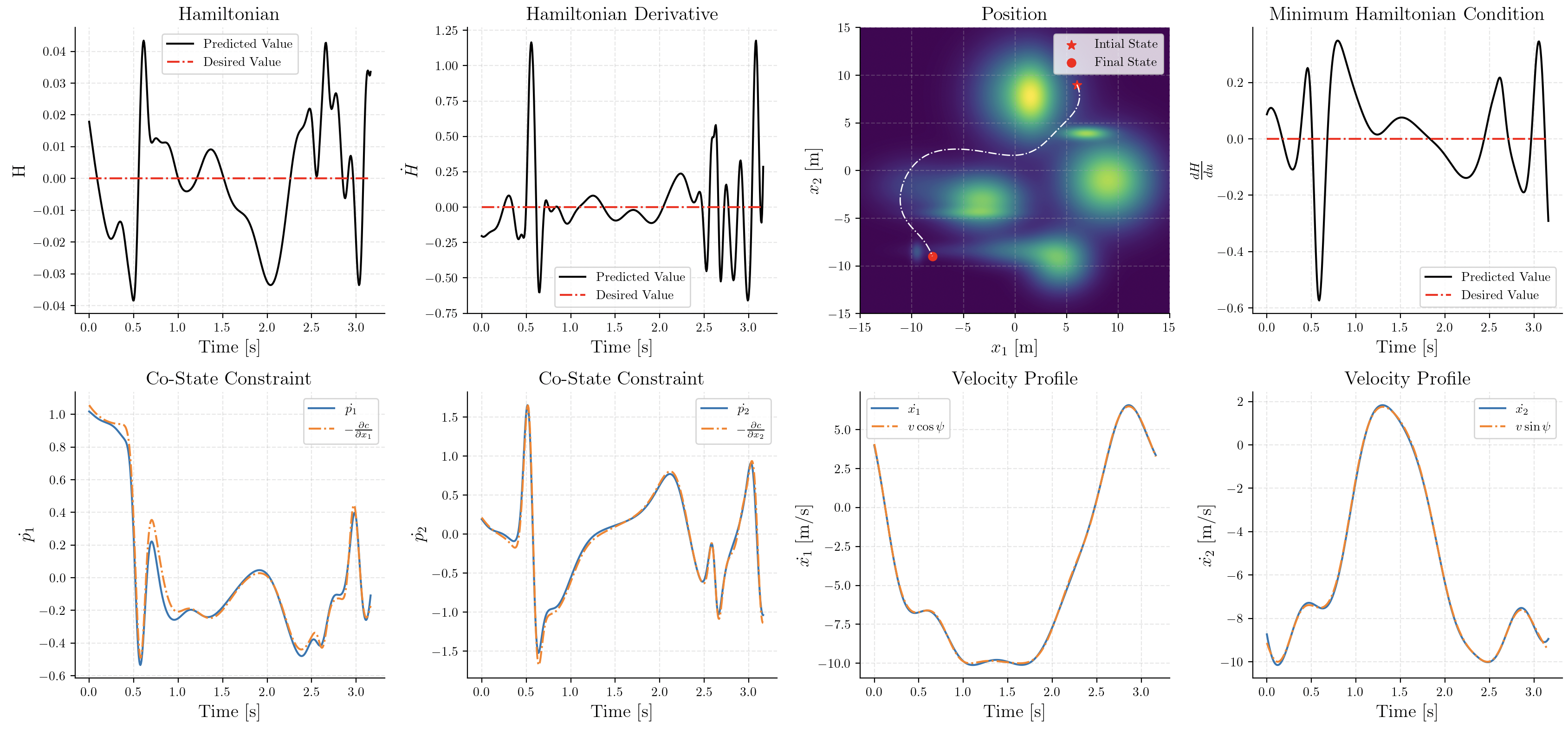}
	\caption{
		A sample PINN output for a single path between two fixed end points
		in a time-invariant threat field. Each plot compares the constraints to the 
		calculated values according the PINN output. 
	}
	\label{fig:single-static}
\end{figure*}

\subsection{Tuning Hyperparameters}

Like any other NN, it is necessary to tune the network
hyperparameters to achieve convergence. We present the best hyperparameters here. 
To predict $\hat{\x},$ $\hat{\psi}$, and $\htf$, we use 
a neural network with 3 hidden layers of 128 neurons and adaptive sine  
activation functions. To predict $\hat{\p}$, we use a neural 
network with 5 hidden layers of 128 neurons and sigmoid linear unit (SiLU) 
activation functions. We choose to separate these networks as the sine activation 
function performs well to predict the states whereas the SiLU provides a good representation of the co-states.
Note that while there are 2 distinct sets of weights and biases for the two networks, we pass
all parameters to the same optimizer and perform gradient updates for both networks simultaneously.

We implement the NN using the PyTorch software library. All
required derivatives are calculated
by automatic differentiation using the
{\small \texttt{torch.autograd.grad}} tool. We use the adaptive 
moment estimation (``adam'') optimization method,
with an initial learning rate of $10^{-3}$. We decay the learning rate by a 
factor of 2 (for static fields) and 10 (for time-varying fields) after 7,500 epochs if the network has not yet converged. We run
each training iteration for a maximum of 10,000 epochs and specify 
stopping criteria based on a minimum desired accuracy.

For each hidden layer of the state network, we use adaptive sine as the activation
function due to the sine and cosine terms in the system dynamics
\eqnnt{eq-pmp-state}. Adaptive activation functions have been shown to help the 
network converge faster with minimal addition of parameters 
\cite{jagtap_adaptive_2020}. The general expression for
the adaptive activation function is $\alpha \sin(\beta x),$
where $\alpha$ and $\beta$ are tunable parameters and are passed to the optimizer 
similar to the weights and biases.

Due to the large number of parameters in the co-state network, we use L2 regularization for the co-state network weights to minimize the chance of overfitting, which adds a penalty term to the loss function proportional to the squared magnitude of the model's parameters. It promotes smoother, more stable solutions. The appropriate loss term is $\Loss_{reg}=\lambda |\theta|^2$, where lambda controls the magnitude of the penalty term. We take $\lambda=10^{-6}$.

We use weights to prioritize loss functions which are more difficult
for the network to represent, such as the Hamiltonian and the
co-states. Note that $w_{\Loss_2}=0$ in the time-varying case, as including this loss term 
was found to degrade performance of the network. In comparison, $\Loss_2$ in the static case improved minimization of the Hamiltonian. 
Table \ref{tab:loss-weights} shows the hand-tuned weights
for each loss function in the static and time-varying cases.

\begin{table}
	\caption{Weights used for different loss function terms.}
	\centering
	\label{tab:loss-weights}
	\begin{tabular}{l|lllllllll}
		\toprule
		& $w_{\Loss_1}$ & $w_{\Loss_2}$ &	$w_{\Loss_3}$ & $w_{\Loss_4}$ & 
		$w_{\Loss_5}$ \\
		Static & 100 &  1 &  1 &  1 &  50 \\
		Time-varying & 100 &  0 &  2 &  2 &  200 \\ \midrule
		& $w_{\Loss_6}$ & $w_{\Loss_7}$ & $w_{\Loss_8}$ & $w_{\Loss_9}$ \\
		Static  &  50 &  1 &  50 &  50 \\
		Time-varying &  200 &  75 &  50 &  50 \\
		\bottomrule
	\end{tabular}
\end{table}

We also implement learning rate annealing to improve convergence. Annealing adds weights to different terms in the loss function based on gradient
statistics. It is particularly useful in multi-objective optimization as it encourages equal 
prioritization of different loss functions and helps avoid over-minimization of a particular loss term. 
The adjustment to the total loss follows a general process:
\begin{align*}
	\hat{\lambda}_k(n) &= \frac{\max |\nabla_\theta\Loss\msub{ref}(n)|} {\overline{|\nabla_\theta\Loss_k (n)|}}, \: k\in \{1, \ldots, 10\}, \\
	\lambda_k &= \alpha \lambda_k(n-1) + (1 - \alpha) \hat{\lambda}_k(n), \\
	\theta^{n+1} &= \theta^n - \eta\nabla_\theta,  
	\biggl( \Loss\msub{ref}(n) + \sum_{k=1}^{10} \lambda_k(n) \Loss_k(n)\biggl),
\end{align*}
where $\Loss\msub{ref}$ is a reference loss, $\theta$ are the parameters of the PINN,
$\overline{|\nabla_\theta\Loss_k (n)|}$ is the mean value of the
gradient w.r.t. $\theta$, $\alpha$ is a 
hyperparameter, and $\eta$ is the learning rate defined for the
network. A recommended value of $\alpha$ is 0.9
\cite{bischof_multi-objective_2021}.

\section{Results and Discussion}
\label{sec-results}

\def\a{\vec{a}}
\def\half{\textstyle{\frac{1}{2}}}

We perform {numerical simulations} in a square workspace $\wsp
= [-15, 15]$ m. The speed of the vehicle is fixed at $\spd = 10$
m/s. We construct a time-invariant threat field as a finite series
of $\nParam \in \nat$ radial basis functions
\begin{align*}
	\hspace{-2ex}
	\threat(\x) &= 1 +  5 \sum_{i=1}^{\nParam} a_{0,i} \exp
	\left( \half (\x - \a_i)^\intercal \Lambda_i (\x - \a_i) \right),
\end{align*}
where $a_{0,i} \in \real,$ $\a_i \in \real[2],$ and $\vec{0} \prec
\Lambda_i \in \real[2 \times 2]$ are constants chosen to represent the
peaks, centers, and spreads  of the radial bases (centers are locations
of locally maximum threats). The proposed PINN is not dependent on the
specific form of the threat field above. Rather, we choose this form
to allow for analytic calculation of the threat gradient.

\subsection{Time-Invariant Threat Field Results}

To learn a path between a fixed initial and final coordinate in a static threat field, we 
train the network on 512 collocation points. \figf{fig:single-static} 
shows the PINN-generated trajectory between two fixed locations. The plots show, in
the top row, from left to right: the value of Hamiltonian over
using PINN-predicted values for the states and
co-states; the derivative of the Hamiltonian computed using {\small
	\texttt{torch.grad.autograd}}; the vehicle trajectory overlaid on the
threat field, where the brighter, yellow colors indicate higher
intensities and darker, blue colors indicate lower intensities;
$\parders{\hamiltonian}{\hdng},$ which should be zero to satisfy
\eqnnt{eq-pmp-control}; and in the bottom row, from left to right: the
co-state and state derivatives w.r.t time, compared compared to the
necessary conditions in \eqnsnt{eq-optimal-state}{eq-optimal-co-state}.

In this sample result, the error between the necessary conditions in 
\eqnsernt{eq-pmp-state}{eq-pmp-co-state} and \eqnnt{eq-min-threat-Htf} of $O(0.01)$. We observe error 
for necessary condition \eqnnt{eq-pmp-control} of $O(0.1)$. Whereas there is room for 
further reduction in error, these results provide a strong preliminary demonstration of the effectiveness of PINNs in variational optimal control. The PINN provides a trajectory through the environment that avoids high intensity threat while satisfying the vehicle kinematics. Note that for the static case, the constraint imposed by \eqnnt{eq-hamiltonian-derivative} is redundant as we are more concerned with the value of the Hamiltonian.

\begin{figure*}
	\centering
	\includegraphics[width=\linewidth]{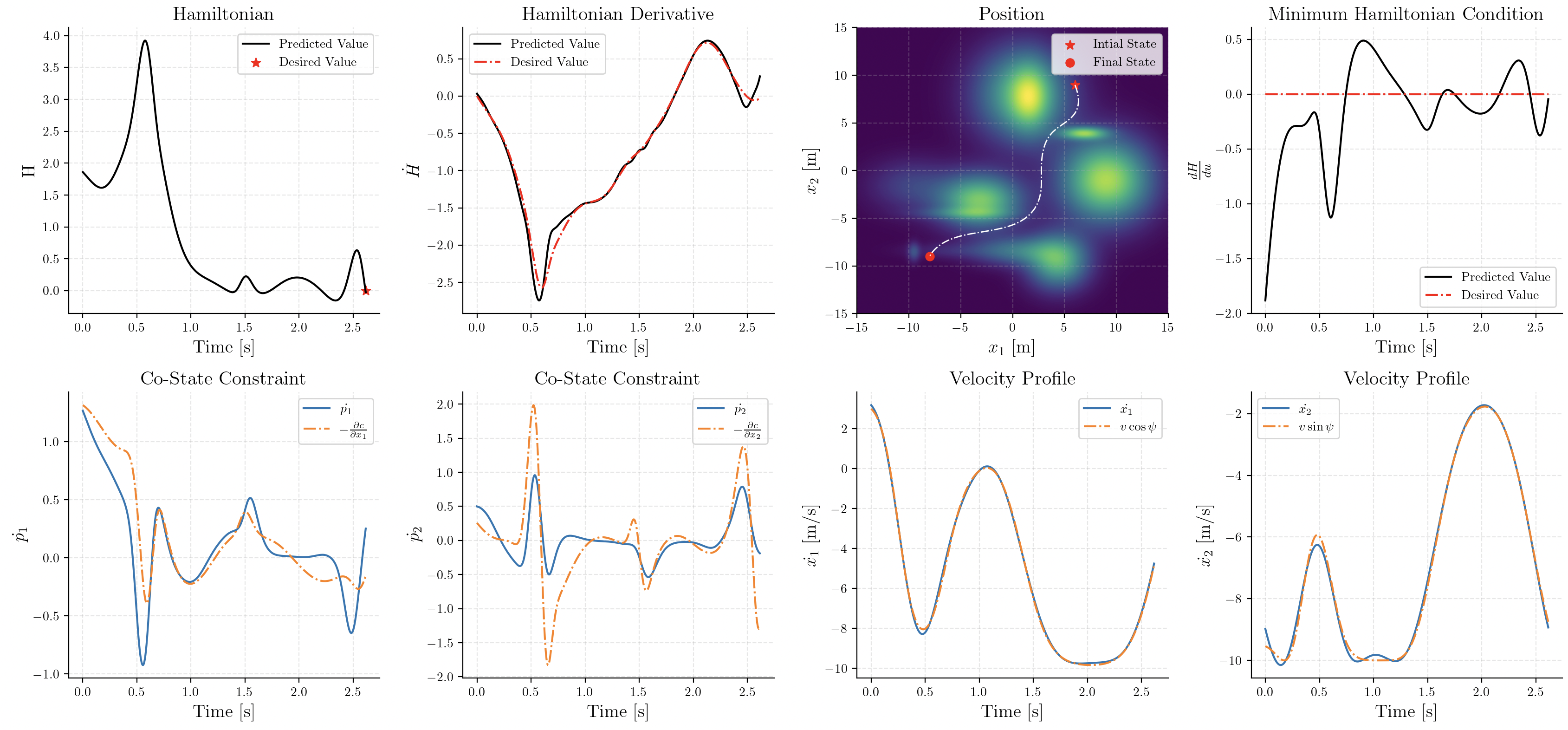}
	\caption{PINN results for a single path between two fixed end points
		in a {time-varying} threat field. Each plot compares the constraints to the 
		calculated values according the PINN output. }
	\label{fig:single-dynamic}
\end{figure*}

\begin{table}
	\caption{Performance metrics for time-invariant threats. $\Loss_{i}$ $\times 10^{-3}.$}
	\centering
	\label{tab:parametric-static}
	\begin{tabular}{l|lllllllll}
		\toprule
		& $\Loss_1$ & $\Loss_2$ &	$\Loss_3$ & $\Loss_4$ & $\Loss_5$ \\
		$\mu$ & 0.629 &  57.13 &  3.314 &  4.296 &  1.222 \\
		$\sigma$ & 1.337 &  128.4 &  6.244 &  4.465 &  2.524 \\ \midrule
		& $\Loss_6$ & $\Loss_7$ & $\Loss_8$ & $\Loss_9$ & $\delta$\\
		$\mu$  &  1.412 &  11.31 &  0.797 &  12.39 & 0.094 \\
		$\sigma$ &  2.973 &  18.12 &  1.129 &  11.42 & 0.289\\
		\bottomrule
	\end{tabular}
\end{table}

Beyond this specific sample result, we evaluate the PINN's performance
on 50 random pairs of initial and final states in
$\wsp$, retraining the PINN for each initial-final state pair. We use a NVIDIA RTX 2070 GPU, and the average training time is 466 seconds.
\tbl{tab:parametric-static} provides the mean value ($\mu$)
and the standard deviation ($\sigma$) of \textit{unweighted} loss functions
$\Loss_1-\Loss_9$. As noted in \scn{sec-method}, these losses indicate
the errors of the PINN-predicted outputs according to the PMP
necessary conditions. \tbl{tab:parametric-static} shows low
average error. We note a high standard deviation for
$\Loss_2$, describing $\dot{\hamiltonian}$, but this loss is
included only to minimize $\Loss_1$.

We also consider the analytic solution. As the threat field is static, we obtain the following expression for $\dot{\psi}$ given $H=0$ and differentiating~\eqnnt{eq-optimal-control} to eliminate the co-states. 
\begin{align}
	\dot{\psi}=\myfracB{\spd (\cos\psi \parder{c}{x_2} - \sin\psi 
	\parder{c}{x_1})}{(\lambda + \threat(\x))}
	\label{eq-heading-angle-derivative}
\end{align}

We use a shooting method to find an initial heading angle $\psi_0$ that satisfies the final boundary conditions $\x_f$ subject to~\eqnnt{eq-kinematics} and~\eqnnt{eq-heading-angle-derivative}. We use the Gekko software library in Python to solve the TPBVP under multiple initial guesses for $\psi_0$. The optimal path is taken to be the path with the least incurred cost if multiple candidate minimums are found. The analytic solution is computed on a M1 Mac and takes between 10 and 600 seconds to converge. Note one path failed to converge. Table~\ref{tab:parametric-static} gives the average difference between the PINN-generated path cost and analytic path cost $\delta=\frac{C_{PINN} - C_{analytic}}{C_{analytic}}$, 
where $C$ is the net path cost approximated by a right Riemann sum, i.e. $C = \htf \sum_{i=2}^N c(\hat{\x}^i, \tau^i) \Delta \tau.$

\subsection{Time-varying Threat Field Results}
We consider a time-varying threat field of the form
\footnotesize\selectfont
\begin{align*}
	\threat(\x) &= 1 +  5 \sum_{i=1}^{\nParam} \frac{a_{0,i}}{2} (1.5 + \cos{a_{0, i}t}) 
	\exp \left( \half (\x - \a_i)^\intercal \Lambda_i (\x - \a_i) \right),
\end{align*}
\normalsize\selectfont
where $a_{0,i} \in \real,$ $\a_i \in \real[2],$ and $\vec{0} \prec
\Lambda_i \in \real[2 \times 2]$ are constants chosen to represent the peaks,
centers, and spreads of the radial bases. The cosine term provides
a periodic change in threat intensity at each point in the environment.

\figf{fig:single-dynamic} shows the results of using the proposed PINN
to generate a trajectory between two fixed locations in this
time-varying threat field. The plots are similar to those in
\fig{fig:single-static}, except we provide two snapshots of the threat field as the agent navigates. Note that the Hamiltonian largely matches the value approximation 
determined using the $\parder{c}{t}$, with errors $O(0.1)$ present in the first 0.5 seconds. 
The PINN 
outputs satisfy the vehicle kinematics and co-state dynamics.

\begin{table}
	\caption{Loss statistics for time-varying threats. All values $\times 10^{-3}.$}
	\label{tab:parametric-time-varying}
	\centering
	\begin{tabular}{l|lllllllll}
		\toprule
		& $\Loss_1$ & $\Loss_2$ &	$\Loss_3$ & $\Loss_4$ & $\Loss_5$ \\
		$\mu$ & 1.006 &  - &  3.287 &  20.81 &  0.927 \\
		$\sigma$ & 3.533 &  - &  4.585 &  81.19 &  2.161 \\ \midrule
		& $\Loss_6$ & $\Loss_7$ & $\Loss_8$ & $\Loss_9$ \\
		$\mu$  &  0.785 &  1.040 &  1.129 &  3.870 \\
		$\sigma$  &  1.935 &  1.779 &  5.339 &  5.617 \\
		\bottomrule
	\end{tabular}
\end{table}

Beyond this sample result, 
we evaluate the PINN's performance on the same time-varying threat field over
50 trials. We randomly choose initial and final states for each trial and retrain the PINN for each initial-final state pair. We use a NVIDIA RTX 2070 GPU with an average training time of 635 seconds. 
\tbl{tab:parametric-time-varying} provides the mean value and 
standard deviation for each of the loss functions $\Loss_1 - \Loss_9.$
With the exception of $\Loss_4$, these errors are of $O(0.001),$ which 
are promising preliminary indicators of a successful solution method for
 the general minimum-exposure problem.


%
%

\subsection{Planning Multiple Paths in a Static Threat Field}

\begin{figure}
	\centering
	\includegraphics[width=\columnwidth]{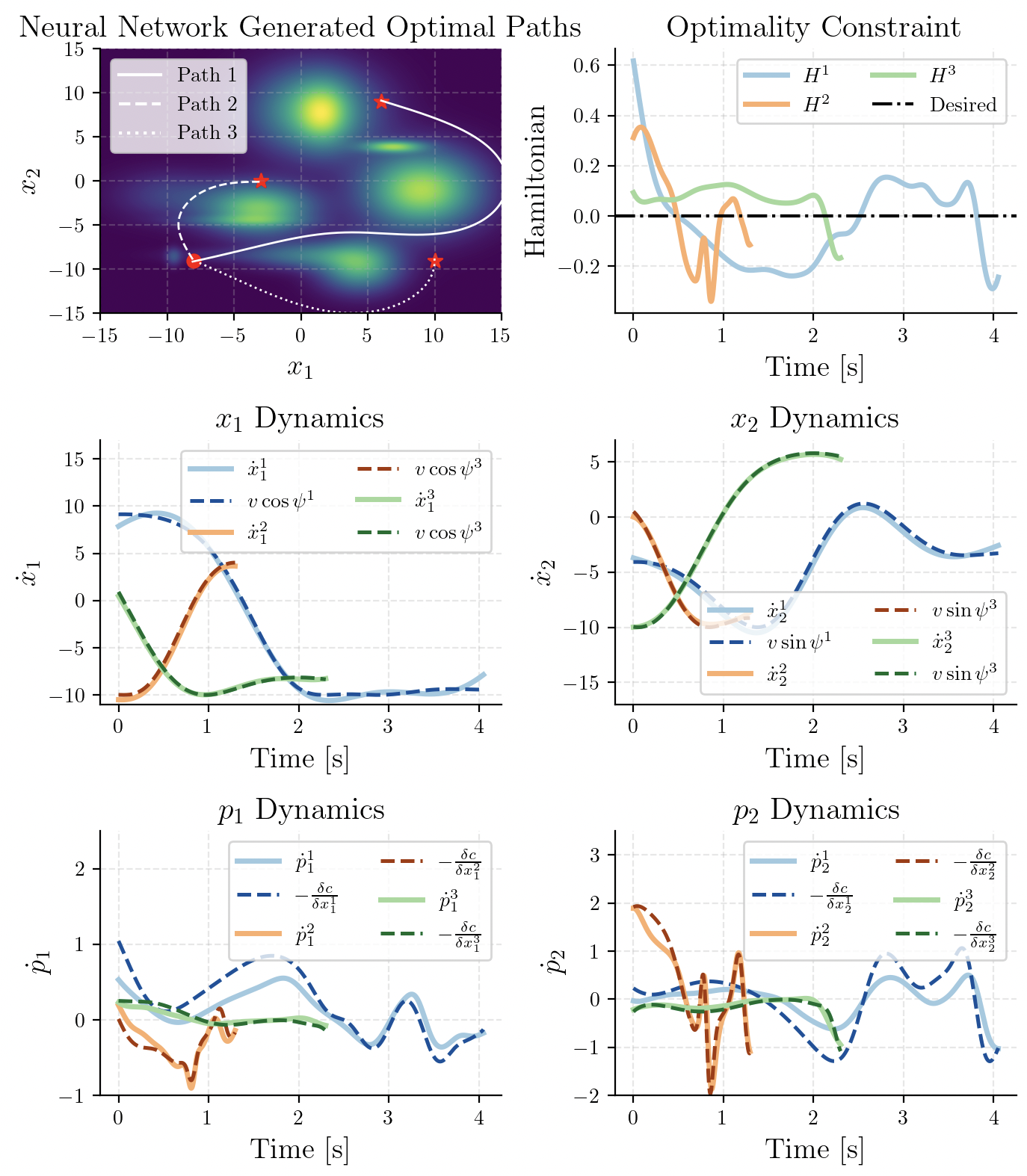}
		
	%
%
%
	\caption{Sample trajectories (upper left), Hamiltonians (upper right), state dynamics (middle), and co-state dynamics (bottom). The superscripts of variables in the legend correspond to the trajectory number (upper left).}
	\label{fig:three-hamiltonian}
\end{figure}

Here we discuss the extension of the PINN to learn optimal
trajectories between \emph{any} pair of initial and final states (as
opposed to a \emph{fixed} pair in the previous two subsections). The
training procedure and loss functions are similar. The main difference
is that, instead of focusing on a fixed trajectory,
we train the network on a large number of initial condition-timestep
pairs for initial and final states in $\wsp.$ Due to the expanded
capabilities of the network, we see a corresponding decrease in
accuracy of the PINN, i.e. larger values reflected in the Hamiltonian.
However, we note that the trajectories accurately capture
the initial and final locations and appear physically reasonable.
\figf{fig:three-hamiltonian} illustrates the primary optimality and 
system constraints for three sample trajectories.


\section{Conclusions and Future Work}
\label{sec-conclusions}

We developed a physics-informed neural network (PINN) to solve the
two-point boundary problem (TPBVP) associated with minimum threat
exposure to a spatiotemporal threat field for a vehicle kinematic
model. This TPBVP, arising from necessary conditions of
Pontryagin's Minimum Principle, is in general difficult
to solve with traditional shooting methods, especially when the threat
is time-varying. The proposed PINN demonstrates a promising new way of
solving this problem. The results presented in this work demonstrate
that the PINN outputs satisfy the necessary conditions with errors of
$O(0.1).$ Based on these promising preliminary results, future work
should be focused on improving the PINN training to further
reduce these errors.


\bibliographystyle{ieeetr}
\bibliography{../references.bib}

\end{document}